\documentclass[aps,prl,twocolumn]{revtex4}
\usepackage[T1]{fontenc}
\usepackage{amsmath}
\usepackage{amssymb}
\usepackage{graphicx}
\usepackage{setspace}
\usepackage{wasysym}
\usepackage{esint}
\usepackage{amsmath,bm}
\usepackage{color}
\usepackage{enumitem}

\usepackage{appendix}

\usepackage{mathtools, nccmath}

\newcommand{\be}{\begin{equation}}
\newcommand{\ee}{\end{equation}}

\newcommand{\bea}{\begin{eqnarray}}
\newcommand{\eea}{\end{eqnarray}}

\renewcommand{\Im}{{\rm \, Im\,}}

\begin{document}

\global\long\def\real#1{\mathbb{R}^{#1}}
\global\long\def\trace{\text{trace}}
\global\long\def\del{\nabla}
\global\long\def\cross{\times}
\global\long\def\diff#1#2{\frac{\partial#1}{\partial#2}}
\global\long\def\rot{\del\cross}
\global\long\def\div{\text{div}}

\title{{Singular measures and information capacity of turbulent cascades}}

\author{Gregory Falkovich and Michal Shavit}
\affiliation{Weizmann Institute of Science, Rehovot 76100 Israel}
 \begin{abstract}
{Is there really such a thing as weak turbulence? Here we analyze turbulence of weakly interacting waves using the tools of information theory. It offers a unique perspective for comparing thermal equilibrium and turbulence: the mutual information between modes is shown to be stationary and small in equilibrium but grows linearly with time in weak turbulence. We trace this growth to the concentration of probability on the resonance surfaces, which can go all the way to a singular measure. The surprising conclusion is that no matter how small is the nonlinearity and how close to Gaussian is the statistics of  any single amplitude, a stationary phase-space measure is far from Gaussian, as manifested by a large relative entropy. Though it might be upsetting to practitioners of weak turbulence approach, this is a rare piece of good news for turbulence modeling: the resolved scales carry significant information about the unresolved scales.
The mutual information between large and small scales is the information capacity of turbulent cascade, setting the limit on the representation of subgrid scales in turbulence modeling.
}

\vskip 0.1truecm

\end{abstract}

\maketitle
\begin{figure*}[!]
	\centering
	\begin{singlespace}
		\includegraphics[scale=0.35]{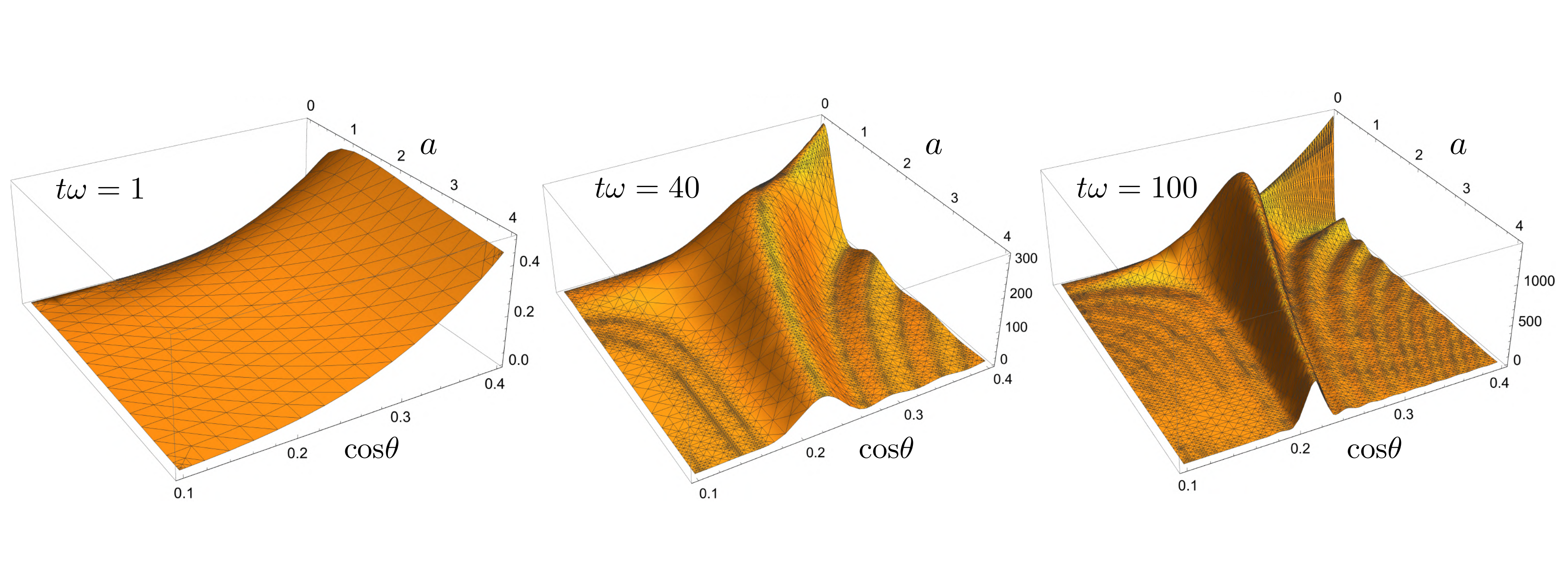}
	\end{singlespace}
	\caption{
		\label{fig:MI} Temporal build-up of the mutual information between three capillary waves in turbulence.
		Here $a=k_2/k_3$,  $\cos\theta_{23}={\bf k_1}\cdot{\bf k_2}/k_1k_2$. For fixed ${\bf k_2}$ and ${\bf k_1}={\bf k_2}+{\bf k_3}$, the resonant surface is the line where the probability and MI peak at $t\omega_2>>1$.
	}
\end{figure*}

There are two quite different perspectives to look at the evolution of a statistical system: fluid mechanics and information theory.
The first one is the continuum viewpoint, where a Hamiltonian evolution of an ensemble  is treated as an incompressible flow in a phase space. Such flows generally mix which leads to a uniform microcanonical equilibrium distribution. On the contrary, to deviate a system from equilibrium, one needs external forces and dissipation that break Hamiltonian conservative nature of evolution and lead to compressible flows in phase space, which generally produce extremely non-uniform 
measures 
\cite{D,EM}.
The second perspective is the discrete viewpoint of information theory, where the evolution towards equilibrium and entropy saturation are described as the loss of all the information except integrals of motion. On the contrary, to keep a system away from equilibrium, we need to act, producing information and decreasing entropy.

Here we make a step in synthesis of the two approaches, asking: what is the informational manifestation of 
nonuniform turbulent measures? Such measures are expected to have a low entropy whose limit is set by an interplay between interaction on the one hand and discreteness, coarse-graining or finite resolution on the other hand.
We shall look at turbulence from the viewpoint of the mutual information (MI), which measures effective correlations between different degrees of freedom.

To keep a system away from equilibrium, environment extracts entropy thus producing information --- where  {is this information encoded?}
Here we consider turbulent systems which  can be treated perturbatively as long as their statistics is close to Gaussian, such as weak wave turbulence (similar approach can be applied to a passive scalar \cite{FGV} and other systems). We show that the {MI between wave modes} is encoded in cumulants 
 (not described by the traditional description in terms of occupation numbers \cite{ZLF}).
The information production builds higher and higher correlations which concentrate sharper and sharper on the resonant surfaces, driving the distribution towards a singular measure.
When nonlinearity is small, we show that the entropy decay is due to the triple moment concentrating  on the three-wave resonance surface, {see Figure 1}.
It is unclear yet how to describe the long-time asymptotic of the entropy decay. When turbulence is driven by a random force, which provides for a phase-space diffusion and smears singularities, the entropy must saturate at a finite value, but the difference with Gaussian random-phase approximation can be large when the Reynolds number is large.

Consider a wave system defined by random complex amplitudes $a_{k}=|a_k|e^{i\phi}$ which satisfy $i\dot{a}_{k}=\delta\mathcal{H}/\delta a_{k}^{*}$ with the Hamiltonian:
\begin{align}
\! {\cal H} =\!\sum_\mathbf{k}\!\omega_{k} |a_{k} |^{2}
+ \sum_{\mathbf{k}\mathbf{p}\mathbf{q}}\!\frac{1}{2} \int \!   \left(V_{kpq}a_{k}^{*}a_{p}a_{q}+\text{c.c.}\right)\delta_{\mathbf{p}+\mathbf{q}}^{\mathbf{k}}\,
 \,.\nonumber
\end{align}
Here  c.c means the complex conjugated terms,
$\delta_{\mathbf{p}+\mathbf{q}}^{\mathbf{k}}$ 
 is the Kronecker delta and we use the shorthand notation $a_k\!\equiv\! a(\bf{k})$, etc. 
The medium is assumed scale invariant, that is both $\omega_k$ and $V$ are homogeneous functions of degree $\alpha$ and $m$, respectively. A central problem is to describe the evolution of the phase-space distribution $\rho\left(\left\{ a_{k},a_{k}^{*}\right\} \right)$. We assume that the second term in the Hamiltonian is on average much smaller than the first one and that the modes are independently distributed at $t=0$. Then the occupation numbers $n_k\delta_{\mathbf{k}'}^{\mathbf{k}}:=\left\langle a_ka_k^{*}\right\rangle$ satisfy a closed kinetic equation \cite{Phase1,ZLF,NR,NJ,Phase2,Mat1,Mat2}:
\begin{eqnarray}
& {d n_k\over dt} =\sum_{{{\bf k}_1{\bf k}_2}}\Im (V_{k12}J_{k12}- 2V_{1k2}^*J^*_{1 k2})\delta^{{\bf k}_1}_{{\bf k}_2+{\bf k}} \,, \label{St0}\\
& J_{123}(t) ={e^{i \omega^1_{2,3}t}-1\over  \omega^1_{2,3}} V_{123}^*(n_2n_3-n_1n_2-n_1n_3)\,.\label{J0}
\end{eqnarray}
The brackets $\left\langle f\right\rangle $ indicate averaging with respect to  $\rho$, $\left\langle a_{i}^{*}a_{j}a_{k}\right\rangle =J_{ijk}\delta_{\mathbf{j+k}}^{\mathbf{i}}$ is the third moment and $\omega^1_{2,3}\!\equiv\!\omega_1\!-\!\omega_2\!-\!\omega_3$. Substituting (\ref{J0}) into (\ref{St0}) gives the collision integral of the kinetic equation, which is a direct analog of the Boltzmann  equation for dilute gases:

\begin{align}&{d n_k\over dt}\equiv St_k=\sum_{{\bf k}_1 {\bf k}_2} \bigl(U_{k12}-2U_{1k2} \bigr)\, \label{St2}\\
&U_{ijk}= {\pi}|V_{ijk}|^2\delta\left(\omega_{jk}^{i}\right)\delta_{\mathbf{j+k}}^{\mathbf{i}}{(n_jn_k-n_in_k-n_in_j)}\,.
\nonumber
\end{align}
The nonlinear interaction time $t_{NL}(k)\!\simeq\!n_k/St_k$ is assumed large relative to the wave period, so that the nonlinearity parameter $\epsilon_k^2\!=\!1/\omega_k t_{NL}(k)$ is small. It is known that the Boltzmann kinetic equation is the first term of a regular cluster expansion only at thermal equilibrium, while even weak non-equilibrium leads to singularities as manifested already in the density expansion of kinetic coefficients like viscosity, diffusivity and thermal conductivity \cite{DC1,DC2,DC3,DC4,Kin1,Kin2}. The kinetic equation for waves describes spectra of developed wave turbulence, both stationary and non-stationary \cite{ZLF,NR,Phase2}, yet the singularities hidden behind this nice picture haven't been analyzed. Here we open this Pandora box and start such an analysis using the most general approach of information theory. That (and processing real data of experiments) requires a discrete approach, so that we consider the number of modes $N$ finite as well as the nonlinearity parameter $\epsilon$. Subtleties related to taking limits $N\to\infty$ and $\epsilon\to0$ are subject of the ongoing work \cite{Phase2,Mat1,Mat2,Phase3,Phase4}.


The self-consistent weak-turbulence description of the one-mode statistics in terms of the occupation numbers $n_{k}$ \cite{Phase1,ZLF,NR,NJ,Phase2,Phase3} {guarantees the statistics of any single amplitude stays close to Gaussian, i.e $q(|a_k|)$ is Rayleigh for every single wave mode.} That tempts one to approximate the whole distribution using only the set of $n_k$, assuming that the amplitudes are independent and the phases are random:
\begin{equation} \label{q}
q\left(\left\{ a_{k},a_{k}^{*}\right\} \right)=\prod_{k}\left(2\pi\right)^{-1}q\left(\left|a_{k}\right|\right),
\end{equation}
which implies a Gaussian approximation for $\rho\left(\left\{ a_{k},a_{k}^{*}\right\} \right)$. Here we show that  $\rho$ is quite different. The difference  between distributions can be measured by the relative entropy (Kullback-Leibler divergence), which is the price of non-optimal coding in information theory: $D\left(\rho\mid q\right)\!\!=\!\!\left\langle \ln\left(\rho/q\right)\right\rangle$. Since $q$ is a product, its entropy is a direct sum of the entropies of non interacting modes: $\sum_\mathbf{k}\ln\left(e\pi n_{k}\right)\!=\!\sum_\mathbf{k}S_{k}$. The relative entropy then coincides with the multi-mode mutual information $D\left(\rho\mid q\right)\!=\!\sum_\mathbf{k}S_{k}-S\left(\rho\right)\!:=\!I\left(\left\{ a_{k},a_{k}^{*}\right\} \right)$. Remind that the mutual information is defined for any subsystems, $A$ and $B$, via their entropies: $I(A,B)=S(A)+S(B)-S(A,B)$.
For example, the mutual information between two parts of the message measures how much of the future part we can predict given the part already received.

Starting with a gaussian $\rho$ at $t\!=\!0$, at the times $1/\omega_{k} \ll t\ll t_{NL}$, the distribution $\rho$ can be determined by the second and third moments using conditional entropy maximum (see Supplementary for details):
\begin{equation} \label{pdf}
 \rho\!=\!\frac{1}{Z}\text{exp}\biggl[-\!\sum_\mathbf{k}\alpha_{k}\left|a_{k}\right|^{2}
	\!+\!\sum_{\mathbf{k}\mathbf{p}\mathbf{q}}F_{kpq}a_{k}^{*}a_{p}a_{q}+\text{c.c}\biggr].
\end{equation}
For $\rho$ to be normalizable, by (\ref{pdf}) we mean a truncated series in powers of $\epsilon$. 
Here we consider terms up to second order. Then  
the parameters $\alpha,F$ of the distribution can be expressed via the moments $J$ and $n$: 
\begin{align}
F_{123}	&=J_{123}^{*}/2n_{1}n_{2}n_{3}\label{F}\\
\alpha_{i}^{-1}	 &=n_{i}-\sum_{\mathbf{k}_{1}\mathbf{k}_{2}}{ \left|J_{i12}\right|^{2}+2\left|J_{12i}\right|^{2}\over 2n_{1}^{2}n_{2}^{2}}.\label{alpha1}
\end{align}
We saw that continuing concentration of the third moment on the resonant surface had no influence on the kinetic equation  (\ref{St2}) since the integral of the imaginary part of the third moment saturates on the short timescale $1/\omega_k$. However, the relative entropy,
\be D(\rho|q)=\sum_{\bold{k}_i \bold{k}_j} {|J_{i+jij}|^2\over 2n_in_jn_{i+j}}, \label{RE2}\ee
is determined by the squared modulus, which depends dramatically on whether the system is in thermal equilibrium or not.
The  equilibrium $n_k=T/\omega_k$ is special because the last bracket in (\ref{J0}) is proportional to $\omega^1_{2,3}$, so the third cumulant is regular everywhere in $k$-space and  constant for long times:
$J_{123}(t) =- V_{123}^*T^2\omega_1/\omega_2\omega_3$.
Therefore, at $t\to\infty$ the relative entropy is small comparing to the total entropy as long as nonlinearity is small:
\be D(\rho|q)=I\{a_k\}=   T {\sum_{ij}{| V_{i+j, ij}|^2 \over   \omega_i\omega_j\omega_{i+j}  } =\left({E_{int}\over T}\right)^2 }\ . \label{RE4}\ee
Away from equilibrium, on the contrary, with time the third cumulant  (\ref{J0}) concentrates in a close vicinity of the resonance surface. It leads to a profound difference between statistics of a wave system in equilibrium and in turbulence. The equilibrium  probability of a configuration $\{a_1,a_2,a_3\}$ is insensitive to resonances, because it is determined  by the instantaneous interaction energy divided by the (uniform) temperature: $exp[-{\cal H}/T]$,  since {$F^*_{123}=J_{123}/2n_1n_2n_3=-V^*_{123}/2T$} in this case. For turbulence, the interaction energy is additionally weighted by the resonance factor $(n_1^{-1}-n_2^{-1}-n_3^{-1})/(\omega_1-\omega_2-\omega_3)$ as the probability is the result of a time averaging. The measure in the phase space is thus regular in equilibrium and tends to singular in turbulence.

The squared cumulants in the relative entropy (\ref{RE4})  have a secular growth in weak turbulence: $\lim_{t\to\infty}|(e^{\imath \Delta t}-1)/\Delta|^2={2\pi}t\delta(\Delta)$. The Liouville theorem requires that this increase of the mutual information and decrease in total entropy is exactly equal to the growth of $S(q)$ due to  the  change in $n_k$ according to
(\ref{St2}):
\begin{align}{dS(q) \over dt}&= \sum_{\bf k}  {1 \over n_k}{dn_k\over dt}=\sum_ {{\bf k}_1 {\bf k}_2 {\bf
k}_3} {1\over 2n_1n_2n_3}{d\over dt}|J_{123}|^2
\ .\label{S2}\end{align}

Hamiltonian evolution by itself does not change the entropy $S(\rho)$, but non-equilibrium state requires pumping and damping by an environment. If its action makes  $n_k$ stationary, then the information production is ultimately due to the entropy extraction by the environment: ${dD(\rho|q)/dt}=- ({dS/dt})_{env}$.
We see that stationarity of the second moment does not mean stationary distribution. On the contrary, the third moment (and other cumulants) are getting more and more singular, reflecting the total entropy decrease and the growth of the relative entropy between the true distribution and the random-phase Gaussian approximation:
\begin{align}
\!\!\!\!D(\rho|q)= t\sum_{{\bf k} {\bf p}  {\bf
		s}} U_{kps}{ n_pn_s-n_kn_p-n_kn_s \over n_kn_pn_s} >0\,.\label{D2}\end{align}
Contribution to the relative entropy of every cumulant is proportional to its square. The fourth cumulant is  $\propto V^2$, so its contribution is proportional to $V^4$ and can be neglected in this order. Formula (\ref{D2})  can be written as
$D=t\sum_kt^{-1}_{NL}(k)$ and {is the first term of the expansion in powers of time, valid at $t<t_{NL}$. The terms with higher powers of time will involve higher cumulants.} One can estimate $t_{NL}^{-1}(k)\simeq \sum_j|V_{k+j,kj}|^2n_j/\omega_j\simeq \omega_k\epsilon_k\propto k^{2m+d-s-\alpha}$.
At $t\simeq t_{NL}(k)$, when nonlinearity at the three-wave resonant surfaces $\omega_j+\omega_k=\omega_{j+k}$ is getting of order unity, the triple moment is expected to stabilize.
At that stage the entropy change already is not small, but could be comparable to the total entropy. At later time, the total entropy decrease is modified, but does not necessarily stop, contrary to what one may suggest. The reason is that the   entropy extraction depends on the environment. We illustrate that for two qualitatively different ways of pumping the system.

Let us first add to the rhs of $\imath{\partial a_k/\partial t}=\partial{\cal H}/\partial a_k^*$ a random force and a damping, $f_k-\gamma_k a_k $,  with  $\langle f_k(0)f_j^*(t)\rangle=\delta_{kj}P_k\delta(t)$. When force and damping are not in detailed balance, i.e $\omega_kP_k/\gamma_k$ is not a constant, we have entropy production:
\be
\left({dS\over dt}\right)_\text{env}\!\!\!\!=2\sum_{\bf k} P_k \int\! \prod_j \frac{da_{j}da_{j}^{*}}{{2 i \rho}} \Bigl|{\partial\rho\over\partial a_k}\Bigr|^2-{2}\sum_{\bf k}\gamma_k\,
.\label{stop1}\ee
which depends on the distribution.  Averaging in this case is over the force statistics. Let us show that if the steady  distribution $\rho$ exists, it must 
have very sharp gradients, proportional to the Reynolds number, so that the entropy is much smaller than $S(q)$.
At the initial perturbative stage, the distribution is given by (\ref{pdf}) and we can substitute (\ref{J0},\ref{F},\ref{alpha1})  into   (\ref{stop1}) and obtain:

	\begin{align}
\sum_{\bf k}\int\! &\prod_j \frac{da_{j}da_{j}^{*}}{{2i}} \rho^{-1}\Bigl|{\partial\rho\over\partial a_k}\Bigr|^2= \sum_{\bf k} \alpha_{k}+O(J^4)
.\label{stop2}\end{align}

Here $\alpha_k$ is given (\ref{alpha1}) where the last two terms are initially  small. The pumping then produces much less entropy than the dissipation region absorbs (any non-equilibrium state consumes information, that is exists between a low-entropy source and a high-entropy sink).  {Indeed, the energy spectral density $\omega_qn_q$ is a decreasing function of $q$ in a direct energy cascade, so  for any $ q>k_{pump}$ we have  $\sum_\mathbf{k}P_{k}n_{k}^{-1}< (\omega_qn_{q})^{-1}\sum_\mathbf{k}\omega_kP_{k}$ and
	\begin{align}\nonumber
\left(\frac{dS}{dt}\right)_{\text{env}}
< 2(\omega_{q}n_{q})^{-1}\sum_\mathbf{k}\left(\omega_{k}P_{k}-\gamma_{k}\omega_{k}n_{k}\right)= 0 \nonumber
	\end{align}
follows from the energy balance $\sum_k \omega_k (P_k-\gamma_k n_k)=0$.}
For a developed turbulence with a wide inertial interval $k_{max}/k_{min}=k_{damp}/k_{pump}=Re\gg1$, the spectrum of the direct cascade is $n_k\propto k^{-s}$, and
the ratio of the negative damping term to the positive pumping term in (\ref{stop1}) can be estimated as
${\omega_{pump}n_{pump} /\omega_{damp}n_{damp}  } \equiv Re^{s-\alpha}\gg1$.
Direct energy cascade requires $s>\alpha$, and indeed the entropy absorption by the small-scale dissipation region by far exceeds the entropy production by the pumping region.
{However, this is only true at the initial perturbative stage.
As time proceeds,
the growth of the cumulants and deviation of distribution from Gaussian decreases $dD/dt$ by increasing the pumping contribution.
For developed turbulence, the gradients $\partial\rho/\partial a_{k}$ in the pumping region must increase by a large factor $n_{pump}/n_{damp}$  to reach the steady measure, which is thus very close to singular.}

Another way of creating non-equilibrium is by adding to the Hamiltonian equations of motion the terms $\gamma_ka_k$, where positive $\gamma_k$ corresponds to an instability and negative to  dissipation. Averaging in this case is over the ensemble of initial conditions. The entropy rate of change $dS_\text{{env}} /dt= 2\sum_{\bf k}\!\gamma_k  \leq 0$ is now independent of the distribution and negative for a steady direct cascade for the same reasons of the energy conservation $\sum_{\bf k}2\omega_k \gamma_k n_k  =0$ and $\omega_kn_k$  being larger in the instability region. That means that  the entropy decreases non-stop and the measure goes all the way to singular unless coarse-graining saturates the entropy decrease. Profound difference between turbulent measures generated by additive force and instability was probably first noticed in \cite{VDF}.

To verify our other predictions, one needs to obtain numerically and experimentally multi-dimensional probability distributions. The simplest is to start from two modes.
The  pair correlation function,   $\left\langle a_{\mathbf{k}}a_{\mathbf{p}}^*\right\rangle=0$ for $\mathbf{k}\neq \mathbf{p} $ due to translation invariance, but the fourth cumulant is generally nonzero and so must be the mutual information (first introduced in \cite{IM} for one-dimensional models of turbulence). In thermal equilibrium and for non-resonant modes in turbulence, steady-state MI must be small for small nonlinearity.
 MI between two modes is given by $I_{\mathbf{k},\mathbf{p}}=|J_{\mathbf{k}, \mathbf{p},\mathbf{k}+\mathbf{p}}|^4/(4n_{\mathbf{k}}n_{\mathbf{p}}n_{\mathbf{k}+\mathbf{p}})^2+|J_{\mathbf{k}, \mathbf{p},\mathbf{k}-\mathbf{p}}|^4/ (4n_{\mathbf{k}}n_{\mathbf{p}}n_{\mathbf{k}-\mathbf{p}})^2\propto \epsilon^4$.
	The $ \epsilon^2$-contribution requires minimum three modes: $I_{{ \bf k},{ \bf p},{ \bf q}}=S(a_k)+S(a_p)+S(a_q)-S(a_k,a_p,a_q)=|J_{{\bf k,p,q}}|^2/2n_kn_pn_q$.
We expect order-unity cumulants (as seen, for instance, in \cite{MIT}) and substantially non-Gaussian  stationary joint distribution for resonant modes in turbulence. Finding that distribution is a well-posed task for a future work.

 Our consideration of the MI growth allows solving
the old puzzle: why the direction of the formation of the turbulent spectrum $n_k\propto k^{-m-d}$ is determined by the energetic capacity? When the total energy  $\sum_{\bf k}\! \omega_kn_k $ diverges at infinity ($m<\alpha$, infinite capacity),  the formation proceeds from large to small scales, that is from pumping to dissipation \cite{FS}. In the opposite finite-capacity case, $m>\alpha$, formation of the cascade was surprisingly found to start from small and proceeds to large scales, {that is opposite to the cascade direction \cite{Galt,NR,CN}. We note that it is the growth of  MI that must determine the direction of the evolution, since it quantifies   the build-up of multi-mode correlations  necessary for a steady non-equilibrium state. For  $n_k\propto k^{-s}=k^{-m-d}$, the growth rate of the three-mode mutual information (learning rate) scales  as $dI_{{ \bf k},{ \bf p},{ \bf q}}/dt\propto k^{2m+d-s-\alpha} =  k^{m-\alpha}$.} One then can characterize the directionality of the information transfer by the sign of $m-\alpha$ --- when it is positive, correlations must be established first at  small scales and then propagate to larger scales.  Since the energetic capacity is also finite for the Kolmogorov spectrum of the incompressible turbulence, tantalizing question is whether it is also formed starting from small scales.
Note that we characterized evolution by the growth rate of MI, which is to be distinguished from the transfer entropy \cite{TE}, which characterizes cause-effect relationships in a steady state.
Remark that though the MI between non-interacting Gaussian wave modes is zero, the multi MI between points in physical space,
$I(x_1,\ldots,x_N)= N \ln\Bigl( {e}\pi N^{-1}\sum_{k=1}^Nn_k\Bigr)-\sum_k\ln( {e}\pi n_k)$,  is positive
whenever $n_k$ are not all the same.

Let us briefly compare our findings with similar effects near thermal equilibrium.
Non-equilibrium anomalies in cumulants were noticed  first in the simplest case of the linear response of a dilute gas. Series in powers of density for viscosity and diffusivity contain infinities  due to Dorfman-Cohen  memory effects in multiple collisions \cite{DC1,DC2,DC3,Pic}.
Another profound difference
is in the two-particle correlation function in two distinct space points at the same time. Outside of the radius of molecular forces, this correlation is zero in thermal equilibrium and nonzero away from it \cite{Kin1,Kin2}. Non-equilibrium build-up of long spatio-temporal correlation is a counterpart to our spectral singularities.

Another analogy worth exploring is with many-body localization \cite{MBL}, where phase correlations  prevent thermalization and keep the system in a low-entropy state. There is a vast literature devoted to cumulant anomalies away from equilibrium, see e.g. books \cite{D,EM,DC2,DC3,DC4,Kin1,Kin2} and numerous references there.  We believe that complementarity of information theory and singular measures will lead to a unified approach to these anomalies.

To conclude, we reiterate our main results: the probability distribution of weak wave turbulence is very far from Gaussian, the mutual information is substantial for resonant modes.

We thank A. Zamolodchikov, V. Lebedev, L. Levitov, K. Gawedzki and R. Chetrite for useful discussions. The work was supported by the Scientific Excellence Center and Ariane de Rothschild Women Doctoral Program at WIS, grant  662962 of the Simons  foundation, grant 075-15-2019-1893 by the Russian Ministry of Science,  grants 873028 and 823937 of the EU Horizon 2020 programme, and grants of ISF, BSF and Minerva.


\newpage

\onecolumngrid
\newpage


\appendix*

\part{ Supplementary Material}

	\section{\label{PDF}{\large The Probability Distribution}}
	Here we outline the derivation of the probability distribution $\rho$ for short times (compared to nonlinear time $t_{NL}$) and its entropy.
	
	The probability distribution determined by the second and third moments, $n_k$ and $J_{kpq}$ respectively, assuming maximum conditional entropy, is an extremum of the functional
	\begin{equation}
	Q[\rho]=S(\rho)-\!\sum_\mathbf{k}\!\!\alpha_{k}n_k
	\!+\!\sum_{\mathbf{k},\mathbf{p},\mathbf{q}} \!\! F_{kpq}J_{kpq}+\text{c.c}+\lambda\int\! \prod_j \frac{da_{j}da_{j}^{*}}{{2i}}\rho,
	\end{equation}
	where $\alpha_{k}$ and $ F_{kpq}$ are the corresponding Lagrange multipliers and the last term is a normalization condition. The solution to the extremum problem, apart from the normalization term, is given by
	\begin{equation} \label{pdf}
	\!\!\rho\!=\!Z^{-1}\text{exp}\biggl[-\!\sum_\mathbf{k}\!\!\alpha_{k}\left|a_{k}\right|^{2}
	\!+\!\sum_{\mathbf{k},\mathbf{p},\mathbf{q}}F_{kpq}a_{k}^{*}a_{p}a_{q} +\text{c.c}\biggr].
	\end{equation}
	For $\rho$ to be normalizable, by Eq.~(\ref{pdf}) we mean a truncated series in $\epsilon=|J_{kpq}|^2/n_kn_pn_q\ll1$, assuming weakness of non-Gaussianity:
	\begin{equation} \label{pdfnor}
	\!\!\rho\!=\!Z^{-1}\text{exp}\biggl[-\!\sum_\mathbf{k}\!\!\alpha_{k}\left|a_{k}\right|^{2}\biggr]\biggl(1+\!\sum_{\mathbf{k},\mathbf{p},\mathbf{q}}F_{kpq}a_{k}^{*}a_{p}a_{q} +\text{c.c}+...\biggr).
	\end{equation}
	The first two terms of the entropy of the distribution are then
	\begin{eqnarray}
	S(\rho)= \sum\limits_\mathbf{k} \ln \pi en_k - {\sum\limits_{\mathbf{ij}}{|J_{i+jij}|^2\over 2n_in_jn_{i+j}}}\ ,\label{S1}
	\end{eqnarray}
	which give the relative entropy:
	\begin{equation}
	D(\rho|q)=  {\sum\limits_{\mathbf{ij}}{|J_{i+jij}|^2\over 2n_in_jn_{i+j}}}\ .\label{RE2}
	\end{equation}
	where $q=\prod_k\frac{1}{\pi n_k}e^{-\sum_k|a_k|^2/n_k}$ is the Gaussian approximation.
	
	\section{{\large Entropy Production}}
	Here we outline the derivation of (13) in the main text:
	\begin{equation}\label{FP}
	\sum_{\bf k}\int\! \prod_j \frac{da_{j}da_{j}^{*}}{{2i}} \rho^{-1}\Bigl|{\partial\rho\over\partial a_k}\Bigr|^2= \sum_{\bf k} \alpha_{k}+O(J^4).
	\end{equation}
	Starting with (\ref{pdfnor}), the derivative of $\rho$ with respect to $a_k$ is given by
	\begin{equation}
	\frac{\partial\rho}{\partial a_{k}}	 =\left(-\alpha_{k}a_{k}^{*}+\sum_{ij}^{N}\left(2F_{ijk}a_{i}^{*}a_{j}+F_{kjl}^{*}a_{j}^{*}a_{i}^{*}\right)\right)\rho
	\end{equation}
	so $\forall k$ 
	\begin{align}\nonumber
	\int\!\prod_j \frac{da_{j}da_{j}^{*}}{{2i}} \rho^{-1}\Bigl|{\partial\rho\over\partial a_k}\Bigr|^2=& \alpha_{k}^{2}n_{k}-\alpha_{k}\sum_{ij}^{N}\left(2F_{ijk}^{*}\left\langle a_{i}a_{j}^{*}a_{k}^{*}\right\rangle +F_{kjl}\left\langle a_{j}a_{i}a_{k}^{*}\right\rangle +2F_{ijk}\left\langle a_{i}^{*}a_{j}a_{k}\right\rangle +F_{kji}^{*}\left\langle a_{j}^{*}a_{i}^{*}a_{k}\right\rangle \right)\\
	&+\sum_{ijlm}^{N}\left(4F_{lmk}^{*}F_{ijk}\left\langle a_{i}^{*}a_{j}a_{l}a_{m}^{*}\right\rangle +F_{kji}^{*}F_{kml}\left\langle a_{j}^{*}a_{i}^{*}a_{m}a_{l}\right\rangle \right)\\
	=&\alpha_{k}^{2}n_{k}-2\sum_{ij}^{N}\left(2\left|F_{ijk}\right|^{2}+\left|F_{kji}\right|^{2}\right)n_{i}n_{j}+O\left(F^{3}\right)\\
	=&\alpha_{k}\left(\alpha_{k}n_{k}-\sum_{ij}^{N}\left(\frac{\left|J_{ijk}\right|^{2}}{n_{i}n_{j}n_{k}}+\frac{\left|J_{kij}\right|^{2}}{2n_{i}n_{j}n_{k}}\right)\right)+O\left(F^{3}\right).
	\end{align}
	Using the first terms in the expansion for $\alpha_k$
	\begin{equation}
	\alpha_{k}=\frac{1}{n_{k}}\left(1+\sum_{ji}\left(\frac{\left|J_{kji}\right|^{2}}{2n_{j}n_{i}n_{k}}+\frac{\left|J_{jik}\right|^{2}}{n_{j}n_{i}n_{k}}\right)\right),
	\end{equation}
	yields (\ref{FP}).

	\section{\large Information Capacity of Turbulent Cascades}
	
	This section includes a complementary and broader discussion to the main text regarding the information capacity of turbulent cascades.
	
	The most important practical problem in modeling multi-mode systems, both in and out of equilibrium, is an inability to resolve all scales. It is then imperative to learn how much information about the whole system one can reliably receive from the scales explicitly accounted for. This can be quantified by the mutual information between small and large scales, which we now compute perturbatively for an interactive wave system in and out of equilibrium. Remind that the relative entropy and the mutual information decrease monotonically upon any partial average.
	
	Let us integrate the $N$-wave probability distribution (\ref{pdf}) over the amplitude and phase of the $N$-th harmonic:
	\begin{eqnarray}
	{\rho\left\{ a_{1}\ldots a_{N-1}\right\}=\frac{\pi}{Z_{N}\alpha_{N}}\exp\left[-\sum_{k=1}^{N-1}\!\!\alpha_{k}\left|a_{k}\right|^{2}\!+\!\sum_{ijk}^{N-1}
		\!\!\left(F_{ijk}a_{i}^{*}a_{j}a_{k}\!+\!\text{c.c}\right)\!+\!\sum_{ijmn}^{N-1}\!\!n_{N}F_{Nij}F_{Nmn}^{*}a_{i}a_{j}a_{m}^{*}a_{n}^{*}\right]} .\label{pdf2}\end{eqnarray}
	We see that the pdf of the remaining modes depends on $F_{Nji}=J^*_{Nij}/2n_in_jn_N$, which according to (2) in the main text,
	$J_{123}(t) ={e^{i \omega^1_{2,3}t}-1\over  \omega^1_{2,3}} V_{123}^*(n_2n_3-n_1n_2-n_1n_3)$,
	can be expressed via $n_in_jn_N$ for weakly-interacting wave system. However, the entropy of the remaining modes is independent of the eliminated modes in this order:
	\begin{eqnarray} &S_{N-1}= \sum\limits_k^{N-1} \ln \pi e n_k -\sum\limits_{ijk}^{N-1}{|J_{ijk}|^2\over 2n_in_jn_k}+{\rm O}(\epsilon^4)\ .\label{SN-1}\end{eqnarray}
	As a result, up to $\epsilon^4$-terms, the mutual information between the two sets of waves with the wave numbers below and above some $k$ respectively is determined by the triple correlation between the waves taken from different sets:
	\begin{eqnarray}
	I(k)=S(1,k)+S(k,N)- S_N =\sum_{j<k<i}{|J_{ijl}|^2\over 2n_in_jn_l}\ .\label{I0}\end{eqnarray}
	It can be called the information capacity of the cascade. 
	If $k$ is the resolution scale then $dI(k,t)/dt$ measures the rate of the information loss. More and more correlations appear at finer and finer scales, all  lost below the resolution scale.
	
	Eliminating the $N$-th harmonic thus decreases the mutual information by removing some contributions from the sum of all positive terms, while the changes in the remaining terms appear only in the next order. The expression for the relative entropy, (9) in the main text,
	\begin{equation} D(\rho|q)=I\{a_k\}=   T {\sum_{ij}{| V_{i+j, ij}|^2 \over   \omega_i\omega_j\omega_{i+j}  } =(E_{int}/T)^2 }\ , \label{RE4}\end{equation}
	is uniformly valid for thermal equilibrium in a weakly interacting system of waves.   The same is true for the general perturbative expression (\ref{RE2}) at any given time only as long as  the measure did not deviate far from Gaussian.
	
	\subsection{\large  Scaling of the Relative Entropy and Mutual Information}
	
	It is interesting to find out how the relative entropy (multi-mode mutual information) depends on the number $N$ of the degrees of freedom and compare it with the linear dependence of the extensive entropy $S(q)$. For example, MI  of a system with the pair interaction (spins, neurons) grows with $N$  as the number of pairs, $N(N-1)/2$,   at least for low enough $N$. Finding the critical $N_c$ when the  quadratic part is comparable to the linear one, one can estimate, for instance, the cluster size of strongly correlated neurons in the brain \cite{Neuro}. {For a wave system in a fixed box, we assume $N\propto (k_{max}/k_{min})^d$. Let us keep the box size $L=2\pi/k_{min}$ fixed, so that $N$ changes as $k_{max}^d$. In computing occupation numbers one must assume that $n_k\propto (k_{max}/k)^s$ so that $S(q)=\sum_k\ln(\pi e n_k)$ is indeed extensive (and not $N\ln N$).} In thermal equilibrium, the relative entropy   is determined by the modes with $k\simeq k_{max}$, so that one estimates $D\simeq  N \epsilon^2_{k_{max}}\simeq N |V(k_{max})|^2T/\omega_{max}^3\propto\epsilon^2  N^{  1 +(2m+d-3\alpha)/d}$, where we denoted the system-scale nonlinearity parameter: $\epsilon =\epsilon_k (k/k_{min})^{2\alpha+s-2m-d}=\epsilon_k (k/k_{min})^{3\alpha-2m-d}$.  The critical number of modes, over which perturbation approach fails and waves cannot be considered weakly correlated, then scales with the nonlinearity as $N_c\simeq \epsilon^{d/(3\alpha-2m- d)}$. On the contrary, when the relative entropy is sub-extensive, the larger the number of modes the less correlated they are effectively.  On the Kolmogorov spectrum, $dD/dt\propto k_{max}^{m+d-\alpha}\propto N^{1+ (m-\alpha)/d}$, which is super-extensive  when $ {\alpha<m}$, that is when the interaction energy grows with $k$ faster than the sum of the energies of the separate waves.
	
	The mutual information (\ref{I0}) in thermal equilibrium scales as $I(k)\propto k_{max}^{2m-m_1+d-2\alpha}k^{m_1+d-\alpha}$, when $m_1+d-\alpha>0$. When $2m-m_1>2\alpha$, the mutual information is sub-extensive, which can be thought as an analog of an area law. In turbulence, MI grows according to $dI(k,t)/dt\propto k_{max}^{2m-m_1+d-3\alpha+1}k^{m_1+2\alpha-m-1}\propto N^{1+(2m-m_1-3\alpha+1)/d}$. Note that the spectrum locality requires $m_1+2\alpha-m-1>0$ \cite{ZLF}.
	
	{For reader's convenience, Table 1 presents parameters for different wave systems with resonant three-wave interaction.
		
		\begin{table}[h]
			\begin{centering}
				\begin{tabular}{|c|c|c|c|}
					\hline
					Wave system & $\alpha$ & $m$ & $m_{1}$\tabularnewline
					\hline
					\hline
					3D acoustic waves & $1$ & $3/2$ & $1$\tabularnewline
					\hline
					2D weakly dispersive waves & $1$ & $1$ & $1$\tabularnewline
					\hline
					Capillary waves on deep water & $3/2$ & $9/4$ & $7/2$\tabularnewline
					\hline
					Capillary waves on shallow water & $2$ & $2$ & $0$\tabularnewline
					\hline
				\end{tabular}
				\par\end{centering}
			\caption{Parameters for different wave systems with resonant three-wave interaction.}
			
		\end{table}
		
		The mutual information is parametrically larger in turbulence than in equilibrium.
		Indeed as stated in the main text the resolved scales carry significant information about the unresolved scales. 
		However, 
		if we are only interested in large-scale properties, the ignorance of small scales carries much higher price in turbulence than in equilibrium.
		
		There are two ways to mitigate the effect of that ignorance: parameterize the unresolved scales by a small set of variables and/or modify the equations of motion of the scales resolved. How well this task is accomplished is to be measured by the relative entropy between the true distribution and that obtained from solving the restricted set of equations.
		In particular,  {Computer modeling poses the question: if there is a way to renormalize the Hamiltonian of the modes explicitly computed so that their statistics is faithfully reproduced.}  For weakly interacting waves,  (\ref{pdf2}) suggests the following  renormalization to account  for subgrid modes at $k\geq L$:
		\begin{eqnarray} \delta {\cal H}=\!\sum_{ijmn}  \sum_{k=L}^{N}\! T^k_{ijmn}a_i^*a_j^*a_ma_n \Delta(k-i-j)\Delta(k-m-n) \,.\label{SG1}\end{eqnarray}
		We need to choose $T^k_{ijmn}$ so that the evolution of $L$ modes with the new Hamiltonian approximates the true entropy of   (\ref{pdf2}) up to $\epsilon^4$.
		This is possible in thermal equilibrium by choosing $T^k_{ijmn}=V_{kij}V_{kmn}^*n_k/T=V_{kij}V_{kmn}^*/\omega_k$; essentially the same procedure is integrating out  fast degrees of freedom in effective  quantum field theories. It is straightforward to see that it does not work away from thermal equilibrium.
		Generally, in turbulence subgrid modes not only renormalize Hamiltonian, but  provide also random forces with a non-trivial statistics.
		{More likely, a correct way to account for subgrid modes in turbulence is a direct renormalization of the probability distribution, rather  than Hamiltonian; this will be treated elsewhere.}
		
		\section{\large  Scattering Processes}\label{sec:scat}
		It is straightforward to include the four-wave scattering into the entropic treatment of weak wave turbulence.
		Consider the Hamiltonian
		\begin{equation}
		{\cal H}=\sum_k\omega_{\mathbf{k}}\left|a_{\mathbf{k}}\right|^{2}+\frac{1}{2}\sum_{i+j=m+n} T_{ijmn}a_{i}^{*}a_{j}^{*}a_{m}a_{n}\,,\label{Ham4}\end{equation}
		{where $T_{ijmn}=T^*_{mnij}$}. We define the renormalized frequency $\tilde\omega_k=\omega_k+\sum_iT_{ikik}n_i$, where as before $n_i=\langle |a_i|^2\rangle$. Denote $  \Delta =\tilde\omega_1+\tilde\omega_2-\tilde\omega_3-\tilde\omega_4$. Similarly to three-wave ineraction, we obtain in the first order the cumulant $J_{ijkl}=\langle  a_i^*a_j^*a_ka_l\rangle-2n_in_k\delta_{ik} {\delta_{jl}}=\langle\langle  a_i^*a_j^*a_ka_l\rangle\rangle$:
		\begin{equation} J_{1 2 3 4} =2T_{1234}^{*}n_{1}n_{2}n_{3}n_{4}\left(n_{4}^{-1}+n_{3}^{-1}-n_{2}^{-1}-n_{1}^{-1}\right) \frac{e^{i\Delta t}-1}
		{\Delta }\delta\left(\mathbf{k}_{1}+\mathbf{k}_{2}-\mathbf{k}_{3}-\mathbf{k}_{4}\right)\,,\label{F0}\end{equation}
		which gives linearly decaying entropy for non-equilibrium spectra.  {Note that $J_{ijij}=0$.}
		
		The distribution determined by $n_i$ and $J_{ijkl}$ is again given by the conditional entropy maximum
		under the assumption  $|J_{ijkl}|^2/n_in_jn_kn_l\equiv\varepsilon^2\ll1$:
		\begin{equation} \rho\{a_k\}=Z^{-1} \exp{\Bigl[-\sum_k\alpha_k|a_k|^2+\sum_{ijkl}G_{ijkl}a_i^*a_j^*a_ka_l \Bigr]}\ .\label{pdf5}\end{equation}
		where the overall normalization and second moment are given by
		
		\begin{align}	
		Z&=\bigl(1 +2\sum_{ijkl}{G^2_{ijkl}\over \alpha_i\alpha_j\alpha_k\alpha_l}\bigr) \prod_{l=1}^N{\pi\over\alpha_l},\\
		n_{i}&=\alpha_{i}^{-1}+2\alpha_{i}^{-2}\sum_{jkl}\frac{G_{ijkl}^{2}+G_{jkil}^{2}}{\alpha_{j}\alpha_{k}\alpha_{l}} .\label{alpha5}\end{align}
		In this case, $G_{1234}=J^*_{1234}/4n_1n_2n_3n_4$, and the relative entropy (the multi-mode mutual information) is expressed via the sum of the fourth cumulants:
		\begin{eqnarray} D(\rho|q)=\sum_{ijkl}{J_{ijkl}^2\over 8n_in_jn_kn_l}\ ,\label{scat}\end{eqnarray}
		which also grows linearly with time in weak turbulence. Apart from the energy, scattering processes conserve also the number of waves, which makes possible two-cascade state with a direct cascade of energy and inverse cascade of waves. In an inverse   cascade, occupation numbers at the sink are larger than at the source, so that $ {dS_{env}/ dt}\simeq 2\gamma_k (n_{damp}/n_{pump}-1)\geq0$. That means  that in a weakly nonlinear regime an inverse cascade  absorbs  entropy and generates  information, that is it cannot exist without a direct cascade, which provides overall entropy production. It is straightforward to show that any stationary weak turbulence producing and absorbing both integrals of motion generates entropy. Examples of wave systems with four-wave scattering are given in Table 2.
		
		\begin{table}[h]
			\begin{centering}
				\begin{tabular}{|c|c|c|c|}
					\hline
					Wave system & $d$ & $\alpha$ & $m$\tabularnewline
					\hline
					\hline
					Surface gravity waves & $2$ & $1/2$ & $3$\tabularnewline
					\hline
					plasmons and manons & $3$ & $2$ & $0$ \text{or} $2$\tabularnewline
					\hline
					Elastic waves in thin plates & $2$ & $2$ & $0$\tabularnewline
					\hline
				\end{tabular}
				\par\end{centering}
			\caption{Parameters for different wave systems with four-wave scattering.}
			
		\end{table}

		One can also consider the general quantum kinetic equation for scattering \cite{ZLF}:
		\begin{eqnarray}\! &\!\!\!\!{\partial n_k\over\partial t}\!=\!\int|T_{k123}|^2F_{k123}\delta({\bf k}\!+\!{\bf k}_1\!-\!{\bf
			k}_2\!-\!{\bf k}_3)\delta(\omega_k\!+\!\omega_1\!-\!\omega_2\!-\!\omega_3)d{\bf k}_1...\! d{\bf k}_4\,,\nonumber\\&
		F_{k123}=(n_k+1)(n_1+1)n_2n_3-n_kn_1(n_2+1)(n_3+1)\,.\label{quantum}\end{eqnarray}
		In the limit $n_k\gg1$ it gives the classical scattering kinetic equation (\ref{F0}). In the opposite limit $n_k\ll1$ it gives Boltzmann equation:
		\begin{equation} {\partial n_p\over\partial t}=\int|T_{p123}|^2(n_pn_1-n_2n_3)\delta({\bf p}+{\bf p}_1-{\bf
			p}_2-{\bf p}_3)\delta(\epsilon_p+\epsilon_1-\epsilon_2-\epsilon_3)d{\bf p}_1 d{\bf p}_2d{\bf p}_3 \ .\label{Boltzmann}\end{equation}

		\section{\large  Directions of Future Studies}

		Let us briefly discuss how one can use the information theory to learn more on different cases of turbulence. Traditionally, turbulence studies have focused on single-point probability distributions and correlations between two or three points. Just like the progress in stochastic thermodynamics and data processing required passing from correlation functions to entropy and mutual information, the future progress of turbulence studies may lie in this direction.  The whole multi-point probability functional, of course, contains all the information, but it is too vast. However, the entropy of this distribution must be manageable and may answer questions hitherto unasked. For example, entropy measures the rate with which we acquire extra information upon increasing spatial, temporal or amplitude resolution. In addition, mutual information will let us quantify how much information we can infer from some  spatial, temporal or spectral domains about the other, unknown, domains.
		
		Further,  note that the mutual information between Fourier modes is interesting even when no waves are possible, as in incompressible turbulence. In this case, the energy flux is cubic in velocity, which suggests studying the three-mode mutual information (going beyond the perturbative approach of \cite{Brown}). In the other extreme, shock creation in compressible turbulence imposes multi-mode phase correlations, which would be very interesting to characterize by the respective multi-mode mutual information. Maxima of multi-mode mutual information may also reveal a  connectivity graph of a system with a finite number of interacting modes.
		
		Much remains to be learnt from the passive scalar turbulence \cite{FGV}. Consider the scalar field $\theta({\bf r},t)$ passively transported by an incompressible flow ${\bf v}({\bf r},t)$ and pumped by  $\xi({\bf r},t)$:
		\begin{eqnarray} (\partial_t+v_i\nabla_i)\theta=\xi\ .\label{pas1}
		\end{eqnarray}
		Both velocity and pumping are Gaussian. The passive scalar turbulence  is amenable to analytic  treatment when velocity is either smooth in space (Batchelor case) or rough in time (Kraichnan case) \cite{FGV}. In the Batchelor case, the scalar statistics is close to Gaussian, yet the statistics of scalar gradients is not. We think that singularity of the measure must manifest itself in angular singularities of cumulants near collinear configurations. The only known case is a cusp for four-point cumulant \cite{BCKL}. One may hope that information-theoretical treatment of this case might be possible within a perturbative approach. In the Kraichnan case, the scalar statistics is close to Gaussian in the limits of large space dimensionality and spatially rough velocity. These two limits are amenable to analytic treatment, where the cumulants can be computed perturbatively. That would be of much interest to compute the entropy and the mutual information for the scalar field and see how they depend on the space dimensionality and the velocity roughness.
		
		Another promising direction could be a renormalization-group (RG) analysis of the information content of turbulence, in particular, the insight it gives into irreversibility of RG as the best way to learn by forgetting, see  \cite{Z,H,Zo}.
		As we have seen, in thermal equilibrium the multi-mode MI is proportional to  $\epsilon_k$.
		The mutual information of the wave system decreases monotonically under coarse-graining. However, two other steps of RG, re-scaling and renormalization, may increase MI depending on the $k$-dependence of $\epsilon_k$.   For example, a perturbative analysis for the system where the lowest nonlinearity is four-wave scattering, described in Section~\ref{sec:scat}, is expected to go along the lines of Wilson epsilon-expansion \cite{WK}. In that case, the Gaussian fixed point will be either stable or unstable depending on the scaling  $\epsilon_k\propto k^{\varepsilon}=k^{s+\alpha-m-d}$, that is $\varepsilon$ plays the role of $4-d$ in our case. The  mutual information between modes will then respectively increase/decrease upon RG flow for negative/positive $\varepsilon$.   This is apparently because some of the information about eliminated degrees of freedom is stored in the renormalized values of the cumulants. Therefore, MI cannot universally play the role of C-function that guarantees the irreversibility of RG flow. Finding the proper informational characteristics for different RG schemes remains the task for the future \cite{Apenko}. Mention also the use of MI for identifying the relevant degrees of freedom and executing RG steps by a machine-learning algorithm without any prior knowledge about the system \cite{MIMLRG}.
		RG  analysis  may also illuminate the profound differences between the fixed points of RG: equilibrium versus turbulence and direct versus inverse cascade. In turbulence, $I({ \bf k},{ \bf p},{ \bf q})$ must be vanishingly small away form the resonant surfaces ${ \bf k}={ \bf p}+{ \bf q}$, $\omega_k=\omega_p+\omega_q$. The perturbative consideration demonstrated the growth on the resonant surfaces, but cannot determine where the growth stops and how the resulting mutual information depends on $\epsilon_k$ and the ratios $k/k_{min}$ and $k/k_{max}$. Finding the ultimate mutual information of the invariant measure of weak turbulence remains the task for future, since the weak turbulence fixed point must be very far from Gaussian. Particularly interesting it is to establish how the total entropy of a turbulence system scales with the number of degrees of freedom to see if there an "area law of turbulence".


\end{document}